\documentclass{svproc}

\usepackage{amsmath}
\usepackage{amssymb}

\usepackage{graphicx}

\usepackage{url}

\begin{document}
\mainmatter  

\title{Construction of soliton solutions of the 
	matrix modified Korteweg-de Vries equation}
\titlerunning{Construction of solitons of the matrix modified KdV}

\author{Sandra Carillo\inst{1,2} \and Cornelia Schiebold\inst{3}}
\authorrunning{Sandra Carillo and Cornelia Schiebold} 
\tocauthor{Sandra Carillo and Cornelia Schiebold}

\institute{
	Universit\`a di Roma  \textsc{La Sapienza}, Dip. S.B.A.I., 16 Via A. Scarpa,  Rome, Italy, \\
	\email{Sandra.Carillo@uniroma1.it},\\ WWW home page:
	\texttt{https://www.sbai.uniroma1.it/{\textasciitilde{}}sandra.carillo/index.html}
\and
	I.N.F.N. - Sezione Roma1, Gr. IV - M.M.N.L.P.,  Rome, Italy \\
\and
	Department of Mathematics and Science Education MOD, 
    	Mid Sweden University,  Sundsvall, Sweden, \\
    	\email{cornelia.schiebold@miun.se} 
}

\maketitle

\begin{abstract}
	An explicit solution formula for the matrix modified KdV equation is presented, which comprises 
	the solutions given in \cite{NODYCON}. In fact, the solutions in \cite{NODYCON} are part of a subclass 
	studied in detail by the authors in a forthcoming publication. Here several solutions beyond 
	this subclass are constructed and discussed with respect to qualitative properties. 
\end{abstract}

\section{Introduction} 

The present article is a sequel to \cite{NODYCON}, where a general approach to the solution theory 
of the matrix MKdV is outlined and certain solutions are explicitly constructed. Actually these solutions 
are part of a family for which a complete classification will be given in the forthcoming article \cite{preprint}.

Here the focus is on solutions beyond the setting of \cite{NODYCON,preprint}. 
In a somewhat experimental spirit, we will examine ways to weaken the assumptions in \cite{preprint},
and initialize the study of some novel solution classes. The emphasis does not lie on completeness, 
but on a qualitative study of phenomena, discussed mainly for the first interesting cases.

The result to start from is a general solution formula presented in {Theorem 1}, 
building on work in \cite{JMP2009,JMP2011}. 
Solution formulas of this kind have been studied for a quite a while, see \cite{CSch,Ma,Sakh,Sch18}  
and the references therein. Here the use of B\"acklund techniques should also be mentioned \cite{CD,Gu et al,RSch,RS}.  
Closely related formulas for scalar equations are known to generate very large solution classes, 
roughly speaking the solutions accessible by the standard inverse scattering method \cite{Blohm}. 
As the situation for matrix equations is much less transparent, the case studies made here are also 
meant as a step towards better understanding the range of our methods. Finally we mention some 
alternative approaches to matrix solutions in \cite{Chen et al,Goncharenko,Levi et al,Sch09}.

\section{An explicit solution class of the ${\sf d}\times{\sf d}$-matrix Korteweg-de Vries equation 
	depending on matrix parameters and $N$-solitons}

We start with stating an explicit solution class for the modified Korteweg-de Vries equation
with values in the ${\sf d}\times{\sf d}$-matrices, 
\begin{equation} \label{mkdv}
	V_t =V_{xxx} + 3 \{ V^2, V_x \},
\end{equation}
depending on matrix parameters.

\begin{theorem} \label{class}
	For $N \in \mathbb{N}$, let $k_1,\ldots,k_N$ be complex numbers such that $k_i+k_j \not= 0$ for all $i,j$, 
	and let $B_1,\ldots,B_N$ be arbitrary ${\sf d}\times {\sf d}$-matrices. 
	
	Define the $N{\sf d}\times N{\sf d}$-matrix function $L=L(x,t)$ as block matrix $L = (L_{ij})_{i,j=1}^N$
	with the ${\sf d}\times{\sf d}$-blocks
	\[   L_{ij} = \frac{\ell_i}{k_i+k_j} Bj ,   \]
		where $\ell_i = \ell_i(x,t) = \exp(k_i x + k_i^3 t)$.
		
	Then
	\[   V =  \begin{pmatrix} B_1 & B_2 & \ldots & B_N \end{pmatrix}
				\Big( I_{N\sf d} + L^2 \Big)^{-1} 
			\begin{pmatrix} \ell_1 I_{\sf d} \\ \vdots \\ \ell_N I_{\sf d} \end{pmatrix} 
	 \]
	is a solution of the matrix modified KdV equation \eqref{mkdv} with values in the ${\sf d}\times{\sf d}$-matrices
	on every domain $\Omega$ on which $\det(I_{N{\sf d}}+L^2) \not= 0$.
\end{theorem}

The proof of Theorem \ref{class}, which is based on results in \cite{JMP2009,JMP2011}, 
is provided in \cite{preprint}. Here we focus on applications and discuss 
some interesting examples.

\begin{remark}  \quad 
\begin{enumerate} 
	\item[a)]
	In \cite{JMP2011} it is shown that the solution class for the matrix KdV equation which corresponds 
	to the class in Theorem \ref{class} comprises the $N$-soliton solutions as derived by the inverse 
	scattering method in \cite{Goncharenko}.
	\item[b)] In contrast to \eqref{mkdv}, the non-commutative mKdV in the form
	\[   V_t = V_{xxx} + 3 \bigm(VV^TV_x+V_xV^TV\bigm) = 0 ,  \] 
	(as for example derived from reduction of the non-commutative AKNS system) admits also non-square 
	matrix interpretation. We refer to \cite{Sch18} for a fairly complete asymptotics of 2-solitons in the vector case.
\end{enumerate}
\end{remark}

\section{Explicit solutions}

Motivated by \cite{NODYCON}, the subclass of solutions arising from choosing $k_1,\ldots, k_N \in \mathbb{R}$ 
and $B_1 = \ldots = B_N =: B$ (up to a common real multiple) where both $B$ and its Jordan canonical form are real,
is discussed thoroughly in \cite{preprint}, the main result being a complete classification of this subclass
up to a possible (common) change of coordinates. 
It should be stressed that all solutions in \cite{NODYCON} belong to this subclass.  

In the present section a variety of solutions beyond this case are presented. 

\subsection{The matrix parameter $B$ does not have a real Jordan form} \label{ss ex1}

A prototypical example for a real matrix without real Jordan form in the case ${\sf d}=2$ are rotations.
Consider 
\[   B= \begin{pmatrix} \ \ \frac{1}{\sqrt{2}} & \frac{1}{\sqrt{2}} \\ - \frac{1}{\sqrt{2}} & \frac{1}{\sqrt{2}} \end{pmatrix} , \]
the rotation with the angle $\frac{\pi}{4}$, and let $N=1$. 

The corresponding solution according to Theorem \ref{class} is
\begin{eqnarray*}
	V 	&=& B \Big( I_2 + \big( \frac{1}{2k} \ell B \big)^2 \Big)^{-1} \ell I_2 \\
		&=& 2k\, g B \Big( I_2 + \big( g B \big)^2 \Big)^{-1} ,
\end{eqnarray*}
where $g = \ell/(2k)$.
Since $B^2$ is the clockwise rotation by $\pi/2$, i.e. $B^2 = \begin{pmatrix} 0 & 1 \\ -1 & 0 \end{pmatrix}$, \\[-1ex] 
this is very easily computed explicitly, giving
\begin{eqnarray*}
	V &=& \sqrt{2}k\, g \begin{pmatrix} \ \ 1 & 1 \\ - 1 & 1 \end{pmatrix} 
			 \begin{pmatrix} 1 & g^2 \\  -g^2 & 1 \end{pmatrix}^{-1} 
		\ =\ \sqrt{2}k\, \frac{g}{1+g^4}  \begin{pmatrix} \ \ 1 & 1 \\ - 1 & 1 \end{pmatrix} 
			 \begin{pmatrix} 1 & -g^2 \\  g^2 & 1 \end{pmatrix} \\
		&=& \sqrt{2}k\, \frac{g}{1+g^4}  \begin{pmatrix} \ \ 1+g^2 & 1-g^2 \\ - (1-g^2) & 1+g^2 \end{pmatrix} 
\end{eqnarray*}
Note that this solution is regular and moves, without changing shape, with velocity constant $-k^2$.

\begin{figure}
\begin{center}
	\includegraphics[scale=0.4]{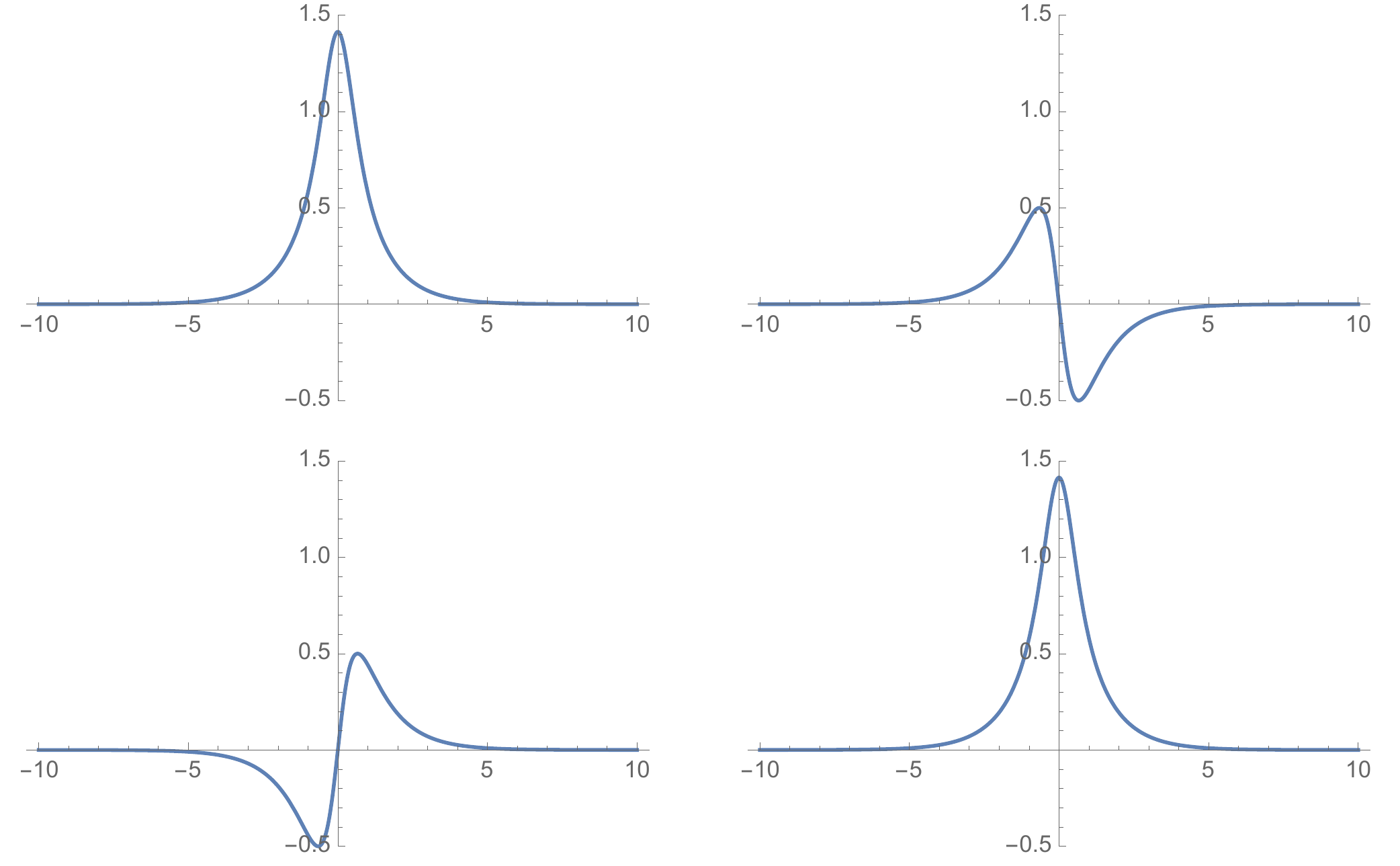}
	\caption{Snapshot of the solution in Subsection \ref{ss ex1} for $k=1$ at $t=0$.}
\end{center}
\end{figure}

\subsection{Unequal matrix parameters} \label{ss ex2}
 
Next we consider the case $N=2$ with matrix parameters $B_1$, $B_2$ such that $B_1 \not= c B_2$
(for all $c \in\mathbb{R}$). 
In this case we get from Theorem \ref{class}
\[   V 
	= \begin{pmatrix} B_1 & B_2 \end{pmatrix} 
	\Big( I_{2{\sf d}}+ M^2  \Big)^{-1} 
	\begin{pmatrix} \ell_1 I_{\sf d} \\ \ell_2 I_{\sf d} \end{pmatrix}    
	\mbox{ with } 
     M
     	= \begin{pmatrix} \frac{1}{2k_1}\ \ \ell_1B_1 & \frac{1}{k_1+k_2}\ell_1B_2 \\ 
	\	\frac{1}{k_1+k_2}\ell_2B_1 & \frac{1}{2k_2}\ \ \ell_2B_2 \end{pmatrix} . 
\]
Let us first focus on the case that \[ B_1 B_2 = 0 = B_2 B_1. \]
Using this assumption, it is straightforward to verify that
\begin{eqnarray}
	\begin{pmatrix} B_1 & B_2 \end{pmatrix} M &=& \begin{pmatrix} B_1 & B_2 \end{pmatrix} R , 
		\label{h1} \\ \label{h2}
	M^2 &=& MR ,
\end{eqnarray}
where
\[   R = \begin{pmatrix} 
		\frac{1}{2k_1} \ell_1 B_1 &0 \\ 
		0 & \frac{1}{2k_2} \ell_2 B_2 
	\end{pmatrix} .
\]
From \eqref{h2} we get $M^3 R = M M^2 R = M (MR)R = M^2 R^2$. Thus, 
$M^2(I_{2\sf d}+R^2) = (I_{2\sf d}+M^2)MR$, showing 
	$(I_{2\sf d}+M^2)^{-1}M^2 = MR(I_{2\sf d} + R^2)^{-1}$, and hence
\begin{eqnarray*}
	(I_{2\sf d}+M^2)^{-1} &=& I_{2\sf d} - (I_{2\sf d}+M^2)^{-1}M^2
		\ = \  I_{2\sf d} - MR (I_{2\sf d}+R^2)^{-1} .
\end{eqnarray*}
Together with \eqref{h1}, this implies
\begin{eqnarray*} \lefteqn{
	\begin{pmatrix} B_1 & B_2 \end{pmatrix} \big( I_{2{\sf d}}+ M^2  \big)^{-1} 
	\ =\ \begin{pmatrix} B_1 & B_2 \end{pmatrix} \Big( I_{2\sf d} - MR (I_{2\sf d}+R^2)^{-1} \Big) }\\
	&=& \begin{pmatrix} B_1 & B_2 \end{pmatrix} \Big( I_{2\sf d} - R^2 (I_{2\sf d}+R^2)^{-1} \Big)
	= \begin{pmatrix} B_1 & B_2 \end{pmatrix} \big(I_{2\sf d}+R^2\big)^{-1}
\end{eqnarray*}
As a result,
\begin{eqnarray*}
	V &=& \begin{pmatrix} B_1 & B_2 \end{pmatrix} 
			\begin{pmatrix} 
				\Big(I_{\sf d}+ \frac{1}{(2k_1)^2} \ell_1^2 B_1^2\Big)^{-1}& 0 \\ 
				0 & \Big(I_{\sf d}+\frac{1}{(2k_1)^2} \ell_1^2 B_1^2\Big)^{-1}
			\end{pmatrix}
		\begin{pmatrix} \ell_1 I_{\sf d} \\ \ell_2 I_{\sf d} \end{pmatrix}    \\
	   &=& \sum_{j=1,2} \ell_j B_j \Big(I_{\sf d}+ \frac{1}{(2k_j)^2} \ell_j^2 B_j^2 \Big)^{-1} \\
	   &=:& \sum_{j=1,2} V_j.
\end{eqnarray*}
Observe that $V_j$ is precisely the solution one obtains from the input data $N=1$ with parameters 
$k_j$, $B_j$ in Theorem \ref{class}. In this sense, $V_j$ can be interpreted as a matrix 1-soliton. 
In the case $B_1B_2 = 0 = B_2B_1$, the solution $V$ therefore is a \emph{linear} superposition 
of the two matrix 1-solitons. 

\begin{figure}
\begin{center} \quad \\[-0.5cm]
	\includegraphics[scale=0.75]{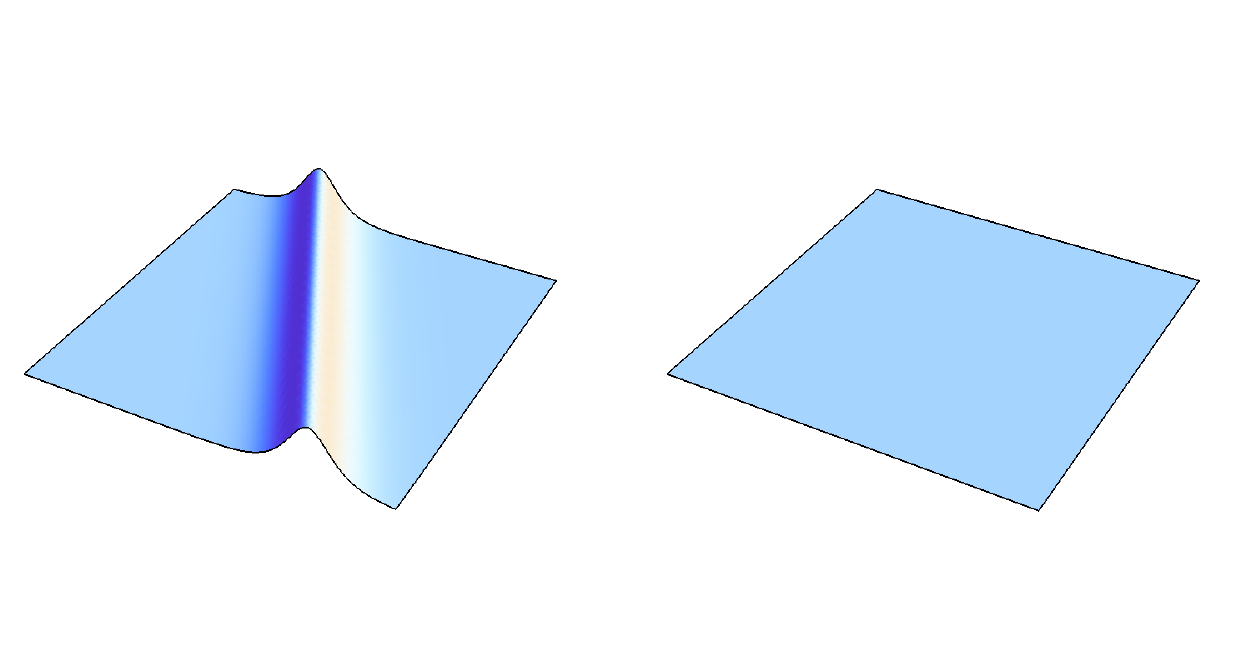} \\[-8ex]
	\includegraphics[scale=0.75]{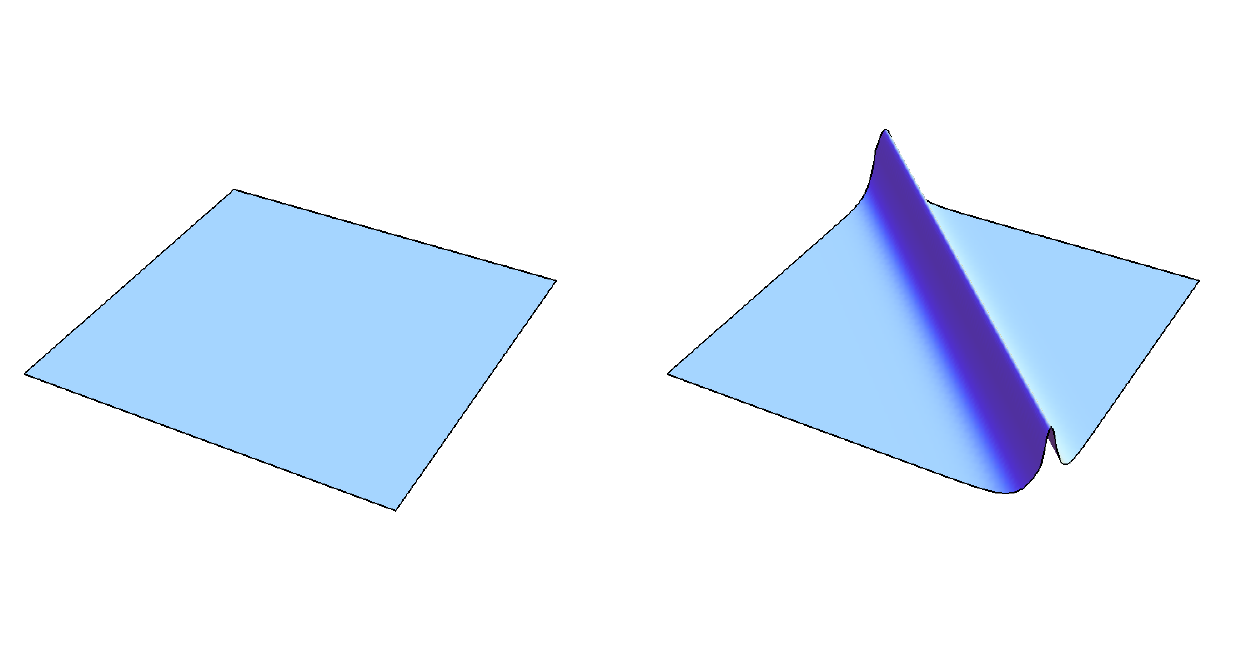} \\[-8ex]
	\caption{The solution in Example \ref{linear ex}a) depicted for 
	$-10 \leq x \leq 10$ and $-5\leq t\leq 5$ with plot range between $-\sqrt{2}$ and $\sqrt{2}$.}
\end{center}
\end{figure}

\begin{figure}
\begin{center} \quad \\[-1cm]
	\includegraphics[scale=0.75]{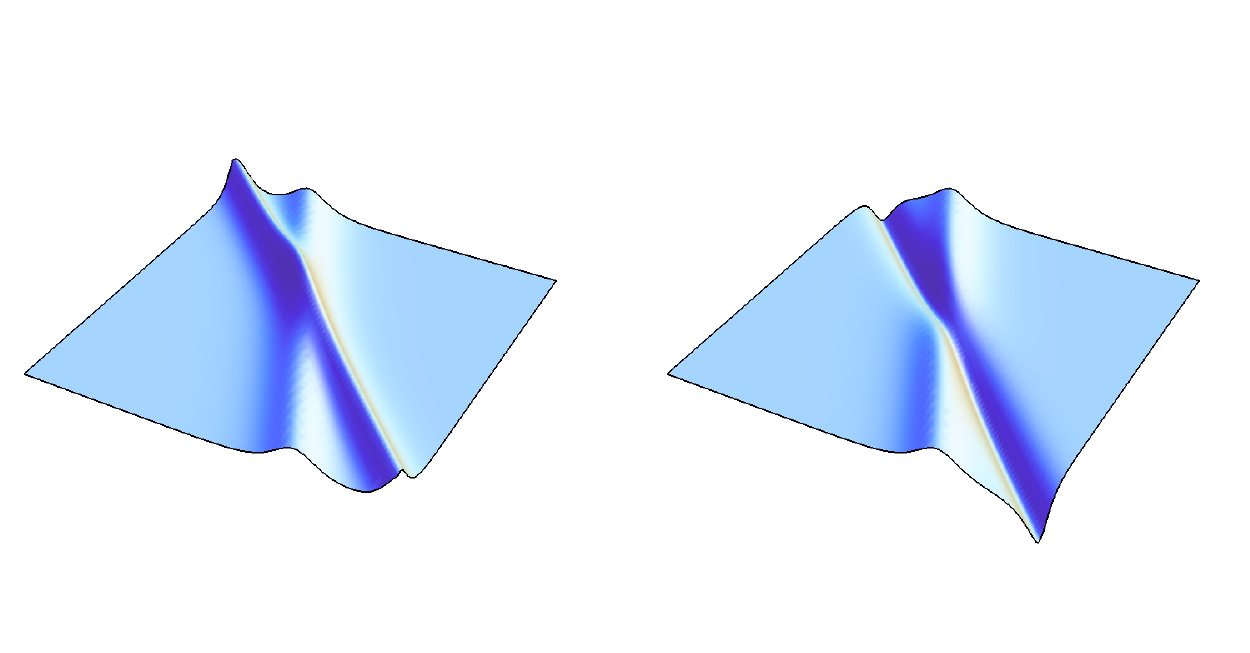} \\[-8ex]
	\includegraphics[scale=0.75]{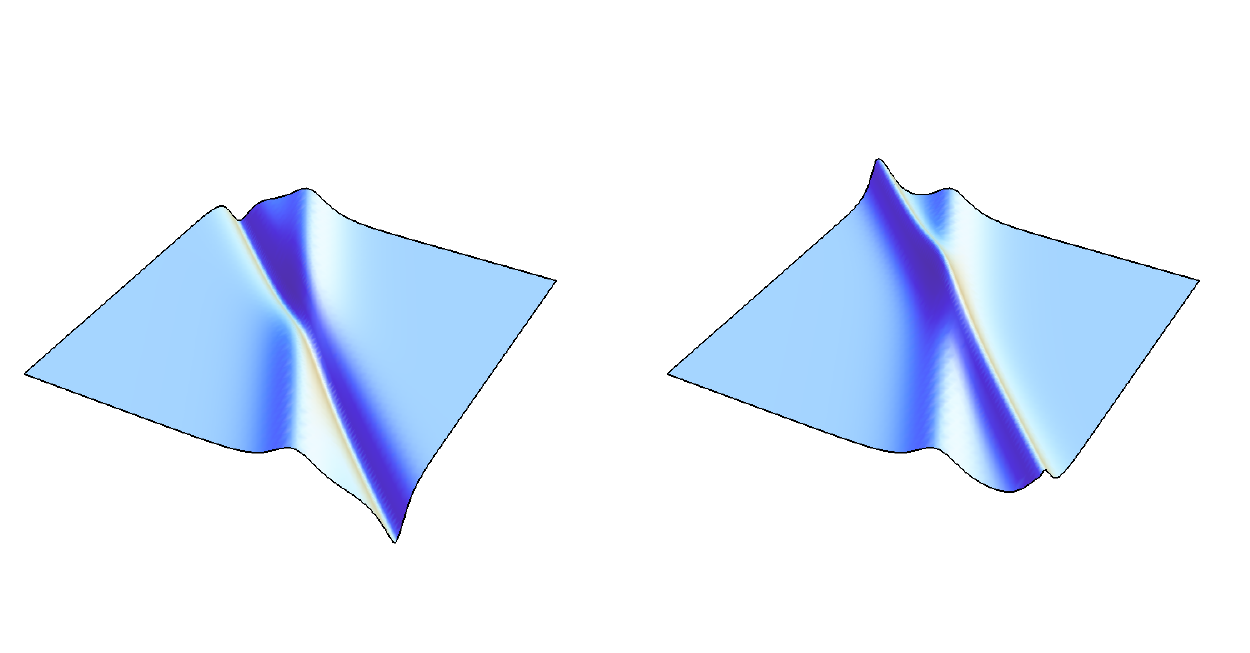} \\[-8ex]
	\caption{The solution in Example \ref{linear ex}b) depicted for 
	$-10 \leq x \leq 10$ and $-5\leq t\leq 5$ with plot range between $-\sqrt{2}$ and $\sqrt{2}$.}
\end{center}
\end{figure}

\begin{example} \label{linear ex}
In Figures 2 and 3, the solution is depicted in the case ${\sf d}=2$, for $k_1=1$, $k_2=\sqrt{2}$, 
and  the matrix parameters are
\begin{description}
	\item[a) {\sc Figure 2}] \quad $B_1 = \begin{pmatrix} 1 & 0 \\ 0 & 0 \end{pmatrix}$, 
						$B_2 = \begin{pmatrix} 0 & 0 \\ 0 & 1 \end{pmatrix}$,
	\item[b) {\sc Figure 3}] \quad $B_1 = \begin{pmatrix} 1 & 1 \\ 1 & 1 \end{pmatrix}$, 
						$B_2 = \begin{pmatrix} \ 1 & -1 \\ -1 & \ 1 \end{pmatrix}$.
\end{description}
\end{example}

Of course there is a huge variety of solutions not covered by the cases above.
We conclude this subsection with one additional example.

\begin{example} \label{last ex} 
	In Figure 4 the solution is depicted in the case ${\sf d}=2$, for $k_1=1$, $k_2=\sqrt{2}$,
	and with the matrix parameters
	\[  	B_1= \begin{pmatrix} 1 & 0 \\ 0 & 1 \end{pmatrix}, \ 
		B_2 =  \begin{pmatrix} 0 & 1 \\ 1 & 0 \end{pmatrix} . \hspace*{4cm}
	\]
\end{example}

\begin{figure}
\begin{center} \quad \\[-2cm]
	\includegraphics[scale=0.75]{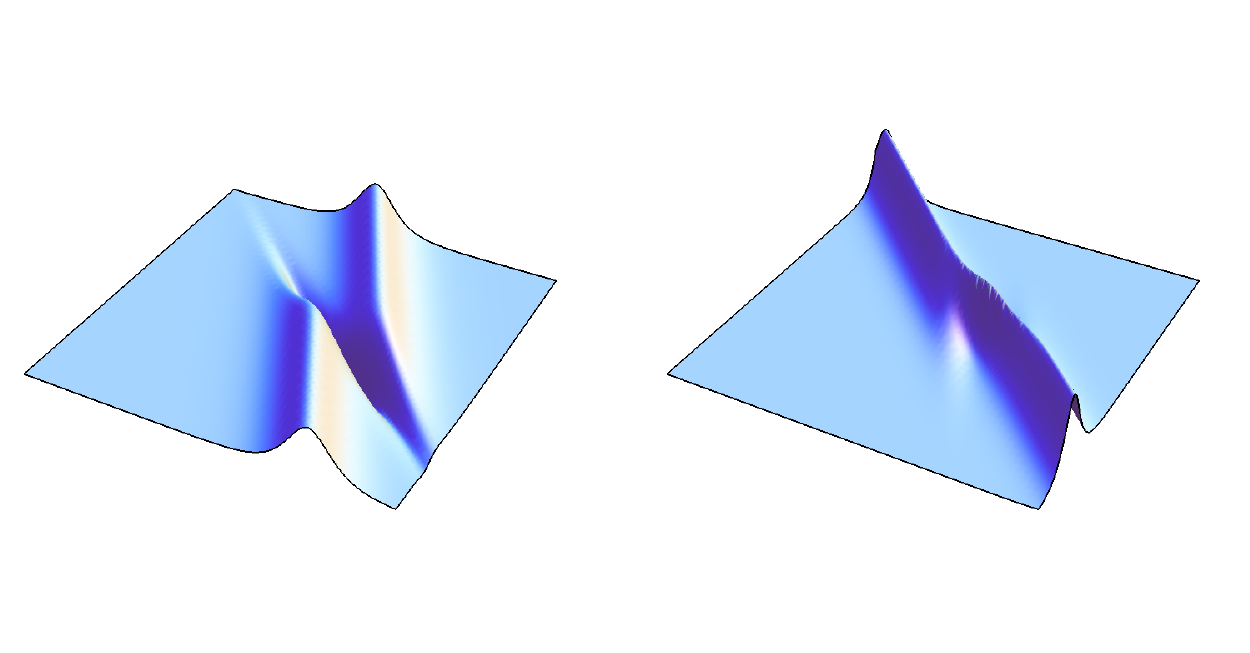} \\[-8ex]
	\includegraphics[scale=0.75]{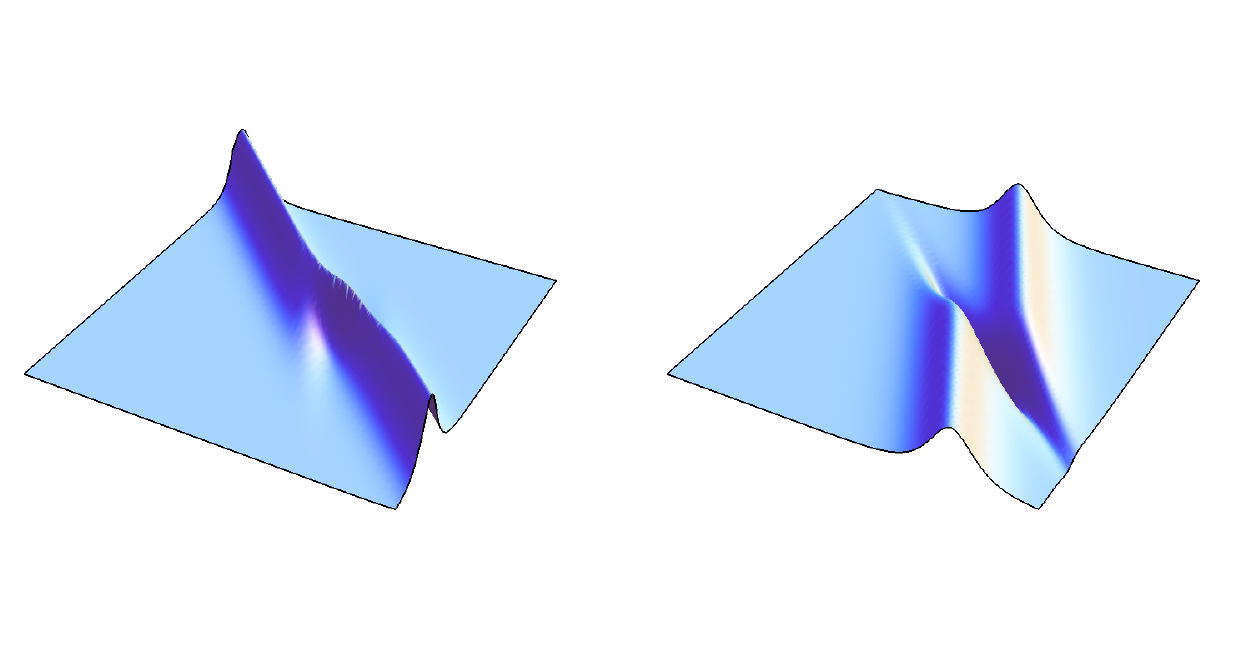} \\[-8ex]
	\caption{The solution in Example \ref{last ex} depicted for 
	$-10 \leq x \leq 10$ and $-5\leq t\leq 5$ with plot range between $-\sqrt{2}$ and $\sqrt{2}$.}
\end{center}
\end{figure}

For comparison, we also depict \\
the \emph{scalar} 2-soliton\footnote{ 
	As generated in the case ${\sf d}=1$ with the input data $N=2$, $k_1=1$, $k_2=\sqrt{2}$, 
	and $b_1=b_2=1$ in Theorem \ref{class}.}.
The frame is \\ the same as in Figure 4.
	
\vspace*{-2cm} \hspace*{7cm} \includegraphics[scale=0.35]{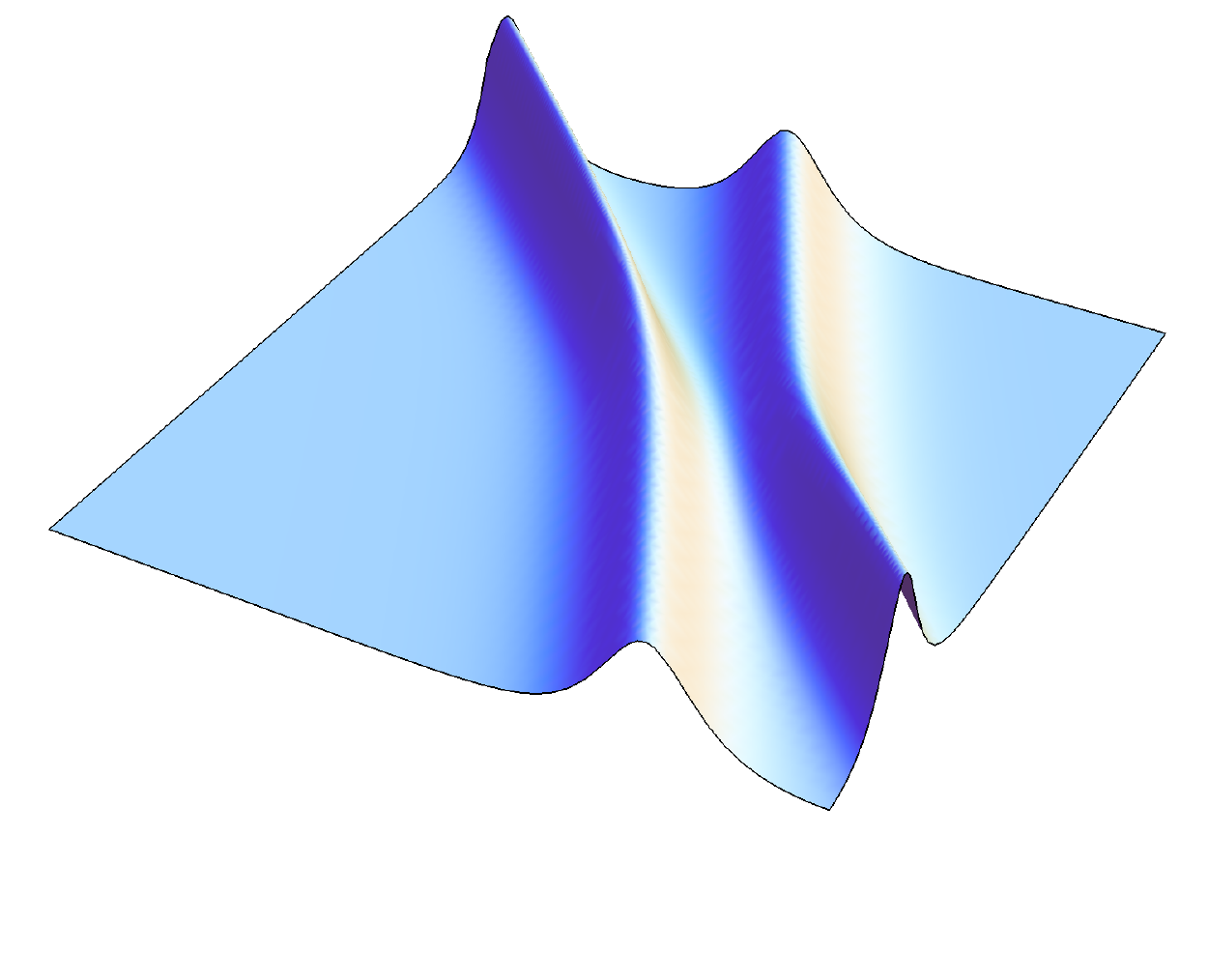} \\[-1.5cm]

\begin{remark}
In \cite{FuCa89}, a B\"acklund chart of KdV-type equations is introduced, linking in particular the KdV 
with its singularity equation (For a generalization of this link to the operator level we refer to \cite{CLPS}, see also \cite{EECT}).   
It is then indicating how this can be used to generate explicit solutions. 
It would be interesting to compare the solution in Example \ref{last ex}, see Figure 4, with 
the (scalar) interacting soliton in \cite{FuCa89}, see Figure 3. 
\end{remark}

\subsection{Complex parameters: A breather solution} \label{ss ex3}

Finally we would like to mention that also complex parameters can lead to real solutions. This is well-known
in the case of the scalar modified KdV equation where the input data $k$, $\overline{k}$, $b$, $\overline{b}$
results in a breather\footnote{
	Here $\overline{k}$ denotes the complex conjugate of $k$.}, 
a solution consisting of a bound state of a soliton and an antisoliton \cite{ZS}.
The same holds true in the case of the matrix modified KdV equation as the following argument shows.

Starting from Theorem \ref{class} with $N=2$ and the parameters chosen to as $k_1=k$, $k_2=\overline{k}$
(such that $\ell_1=\ell$, $\ell_2=\overline{\ell}$) and $B_1=B$, $B_2 = \overline{B}$, our solution reads 
\begin{eqnarray*}
	V &=& \begin{pmatrix} B & \overline{B} \end{pmatrix} 
			 \Big( I_{2\sf d} + M^2  \Big)^{-1} 
			 \begin{pmatrix} \ell I_{\sf d} \\ \overline{\ell} I_{\sf d} \end{pmatrix} 
	 \mbox{ with } 
 	M = \begin{pmatrix} 
			\frac{1}{2k} \ \ell\, B & \frac{1}{k+\overline{k}} \ell\, \overline{B} \\ 
			\frac{1}{k+\overline{k}} \overline{\ell}\, B & \frac{1}{2\overline{k}} \ \overline{\ell}\, \overline{B}
			 		\end{pmatrix}.
\end{eqnarray*}
Introducing $D = \begin{pmatrix} 0 & I_{\sf d} \\ I_{\sf d} & 0 \end{pmatrix}$, such that 
\[	
	\begin{pmatrix} B & \overline{B} \end{pmatrix}  = \begin{pmatrix} \overline{B} & B \end{pmatrix} D
		\quad \mbox{ and } \quad
	\begin{pmatrix} \ell I_{\sf d} \\ \overline{\ell} I_{\sf d} \end{pmatrix} 
			= \begin{pmatrix} \overline{\ell} I_{\sf d} \\ \ell I_{\sf d} \end{pmatrix} 			 
\]
and \\[-6ex]
\begin{eqnarray*}
	DMD  
	&=& \begin{pmatrix} 0 & I_{\sf d} \\ I_{\sf d} & 0 \end{pmatrix}
	   \begin{pmatrix} 
		\frac{1}{2k} \,\ell\, B & \frac{1}{k+\overline{k}} \ell\, \overline{B} \\ 
		\frac{1}{k+\overline{k}} \overline{\ell}\, B & \frac{1}{2\overline{k}} \,\overline{\ell}\, \overline{B}
	   \end{pmatrix}
	  \begin{pmatrix} 0 & I_{\sf d} \\ I_{\sf d} & 0 \end{pmatrix} \\
	  &=& \begin{pmatrix} 
			\frac{1}{2k} \ell\, B & \frac{1}{k+\overline{k}} \ell\, \overline{B} \\ 
			\frac{1}{k+\overline{k}} \overline{\ell}\, B & \frac{1}{2\overline{k}} \overline{\ell}\, \overline{B}
		 \end{pmatrix} .
\end{eqnarray*}
Observe that $D^{-1}=D$. Hence, since $D(I_{2\sf d} + M^2)^{-1}D = (I_{2\sf d} + DM^2D)^{-1}
=(I_{2\sf d} + (DMD)^2)^{-1}$,
we find
\begin{eqnarray*}
	V &=& \begin{pmatrix} \overline{B} & B \end{pmatrix} 
			 \bigg( I_{2\sf d} + 
			 		\begin{pmatrix} 
						\frac{1}{2k} \ell\, B & \frac{1}{k+\overline{k}} \ell\, \overline{B} \\ 
						\frac{1}{k+\overline{k}} \overline{\ell}\, B & \frac{1}{2\overline{k}} \overline{\ell}\, \overline{B}
			 		\end{pmatrix}^2  \bigg)^{-1} 
			 \begin{pmatrix} \ell I_{\sf d} \\ \overline{\ell} I_{\sf d} \end{pmatrix} \ = \ \overline{V} ,
\end{eqnarray*}
showing that the solution $V$ is real.

\begin{example} \label{ex breather}
For illustration, we add two random examples. In both examples $k = 1 + {\rm i}$. For the corresponding scalar breather this implies velocity $=2$, and hence the plots are drawn for $(x,x+2t)$ giving a stationary picture.
The matrix parameter is \\[-2ex]
\begin{description}
	\item[{\sc Figure 6}] \quad
		$B=\begin{pmatrix} {\rm i} & -2 \\ 1+{\rm i} & 2 - {\rm i} \end{pmatrix}$, \\[0.1ex]
	\item[{\sc Figure 7}] \quad
		$B=\begin{pmatrix} {\rm i} & -2{\rm i} \\ 3{\rm i}-1 & -1 \end{pmatrix}$.
\end{description}
\end{example}

\begin{figure} 
\begin{center} \quad \\[-0.5cm]
	\includegraphics[scale=0.75]{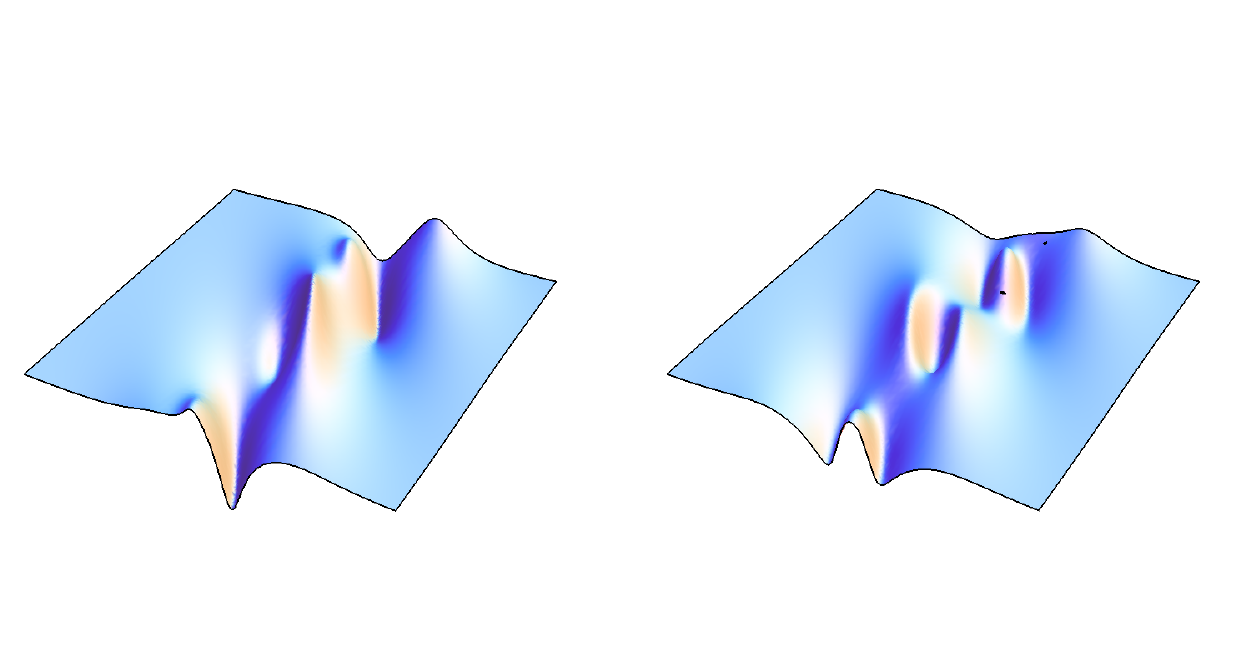} \\[-8ex]
	\includegraphics[scale=0.75]{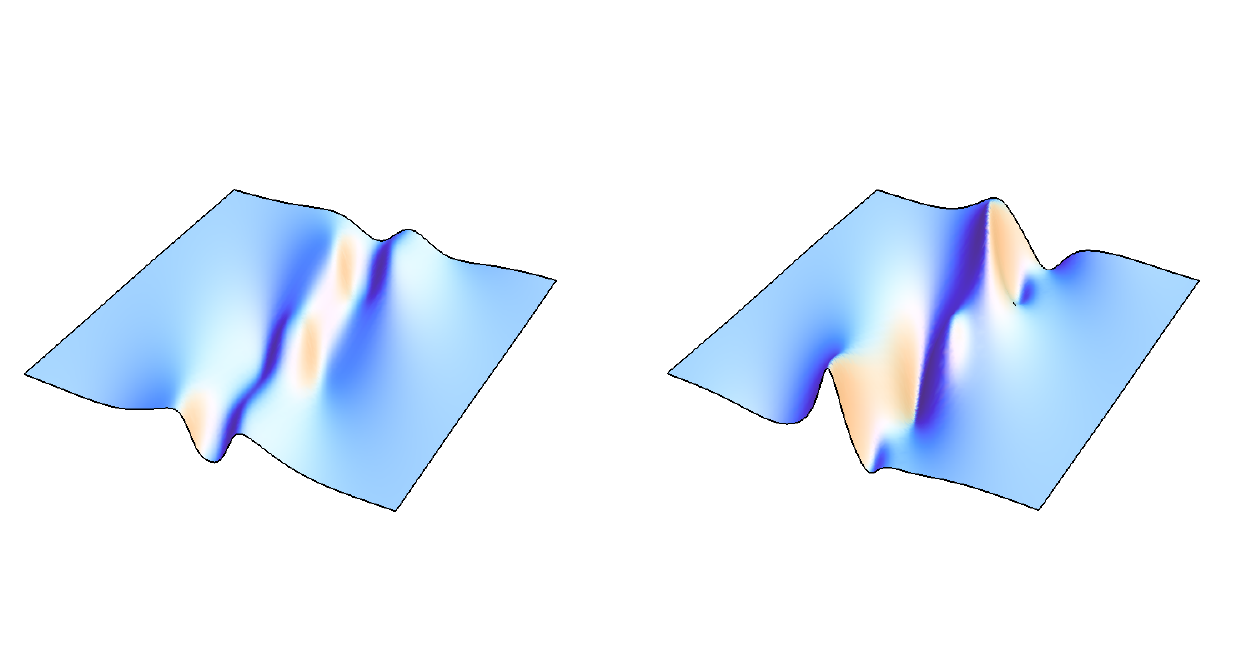} \\[-8ex]
	\caption{The solution in Example \ref{ex breather}a) is depicted for $-5\leq x \leq 5$ and $0\leq t \leq 2$ 
		with plot range between $-3.5$ and $3.5$.}
\end{center}
\end{figure}

\begin{figure} 
\begin{center} \quad \\[-1cm]
	\includegraphics[scale=0.75]{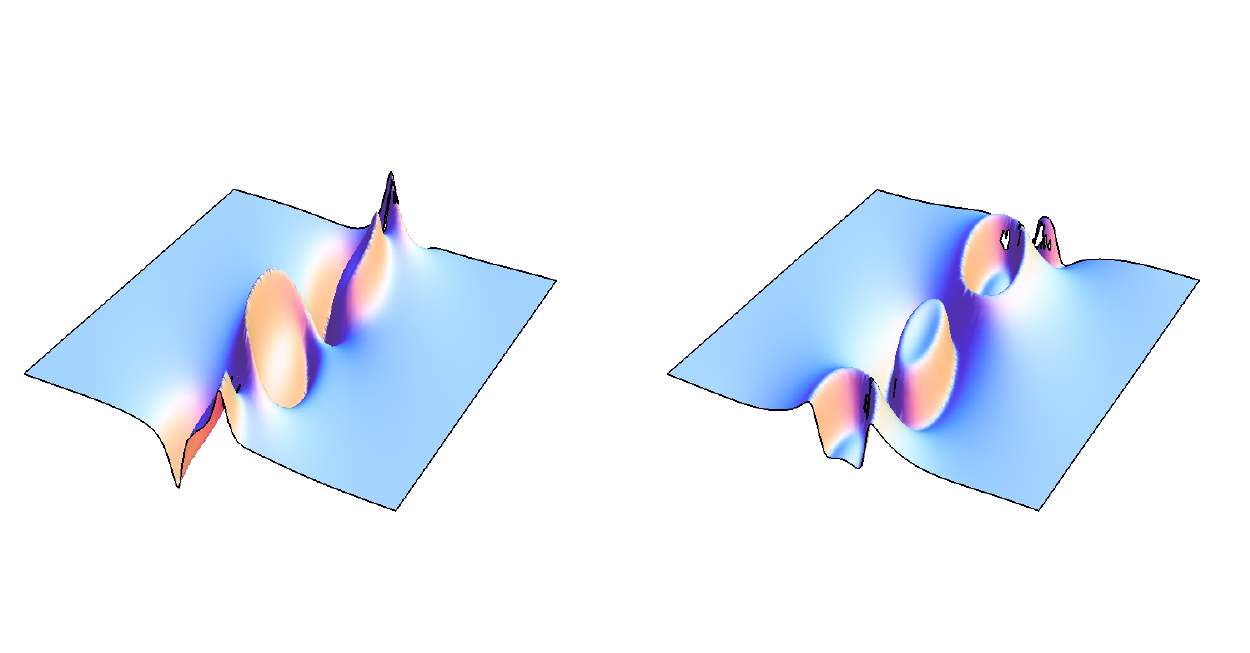} \\[-8ex]
	\includegraphics[scale=0.75]{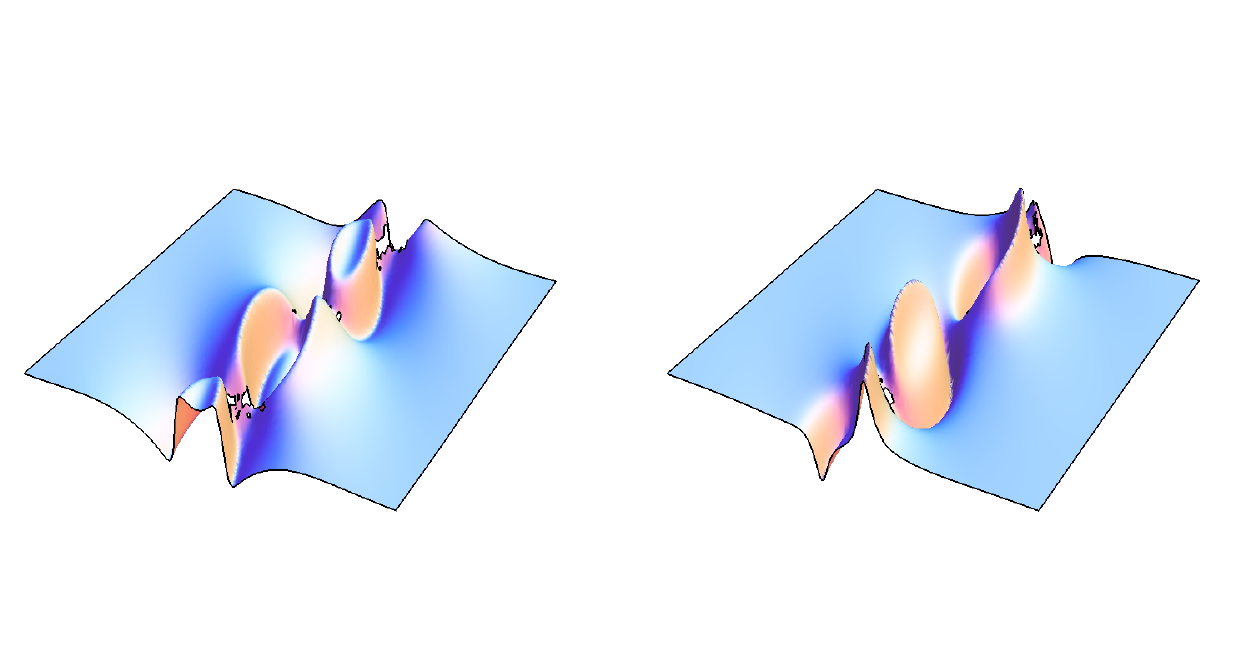} \\[-8ex]	
	\caption{The solution in Example \ref{ex breather}b) is depicted for $-5\leq x \leq 5$ and $-1\leq t \leq 1$ with 
		plot range between $-5.5$ and $5.5$.}
\end{center}
\end{figure}

\subsubsection*{\small Acknowledgements} \small
Under the financial support of G.N.F.M.-I.N.d.A.M.,  I.N.F.N. and Universit\`a di Roma 
 \textsc{La Sapienza}, Rome, Italy.
C. Schiebold acknowledges  Dip. S.B.A.I., Universit\`a di Roma  \textsc{La Sapienza}, for the kind hospitality.

\end{document}